# Electrodynamic modeling of strong coupling between a metasurface and intersubband transitions in quantum wells

Salvatore Campione[1,2,3,#], Alexander Benz[2,3], John F. Klem[2], Michael B. Sinclair[2], Igal Brener[2,3], and Filippo Capolino[1,*]

[1]Department of Electrical Engineering and Computer Science, University of California Irvine, Irvine CA 92697 USA
[2]Sandia National Laboratories, Albuquerque NM 87185 USA
[3]Center for Integrated Nanotechnologies, CINT, Sandia National Laboratories, Albuquerque NM 87185 USA
[#]sncampi@sandia.gov ; [*]f.capolino@uci.edu

*Abstract*—Strong light-matter coupling has recently been demonstrated in sub-wavelength volumes by coupling engineered optical transitions in semiconductor heterostructures (e.g., quantum wells) to metasurface resonances via near fields. It has also been shown that different resonator shapes may lead to different Rabi splittings, though this has not yet been well explained. In this paper, our aim is to understand the correlation between resonator shape and Rabi splitting, and in particular determine and quantify the physical parameters that affect strong coupling by developing an equivalent circuit network model whose elements describe energy and dissipation. Because of the subwavelength dimension of each metasurface element, we resort to the quasi-static (electrostatic) description of the near-field and hence define an equivalent capacitance associated to each dipolar element of a flat metasurface, and we show that this is also able to accurately model the phenomenology involved in strong coupling between the metasurface and the intersubband transitions in quantum wells. We show that the spectral properties and stored energy of a metasurface/quantum-well system obtained using our model are in good agreement with both full-wave simulation and experimental results. We then analyze metasurfaces made of three different resonator geometries and observe that the magnitude of the Rabi splitting increases with the resonator capacitance in agreement with our theory, providing a phenomenological explanation for the resonator shape dependence of the strong coupling process.

*PACS numbers*— 78.67.Pt, 41.20.Cv, 81.07.-b, 78.30.-j.

## I. INTRODUCTION

Metamaterials are artificial materials whose interaction with light results in interesting phenomena including negative refraction [1,2], super-resolution [3,4], artificial magnetism [5], and cloaking [6]. Recently, increasing emphasis has been placed on metasurfaces (MSs), i.e., single layers containing a two-dimensional periodic set of resonators, because their fabrication is compatible with standard semiconductor processing technology and is significantly simpler than that of three-dimensional metamaterials.

Metamaterials are able to localize intense electromagnetic fields in deep sub-wavelength volumes and thus are excellent candidates for the study of strong light-matter coupling – a process in which the excitation energy is periodically exchanged between matter (e.g., quantum wells, QWs) and a metasurface (i.e., a cavity mode). When strong coupling is achieved, the following two properties are simultaneously observed: in the time domain, the electric field radiated by the cavity shows a beating mode at a characteristic time constant; and in the frequency domain, this beating corresponds to a splitting of the single bare cavity resonance into two polariton branches [7,8] with separation $2\Omega_R$, with $\Omega_R$ being the Rabi (angular) frequency.

The first experimental reports of strong coupling using intersubband transitions (ISTs) relied on the use of macroscopic optical cavities [9-16]. Recently, strong coupling has also been demonstrated through the use of a sub-wavelength metal-dielectric-metal microcavity [17] or metasurfaces with subwavelength elements [18-21]. Strong light-matter interaction between planar metasurfaces and ISTs in QWs is easily scalable throughout the infrared spectrum and occurs on a single resonator level involving only a small number of electrons [21]. Previous work using other geometries, for example micro- or nano-cavities, has shown strong coupling at the single resonator level and/or a small number of carriers involved in the process[22-24]. We remark that a metallic backplane is not required to achieve sub-wavelength volumes as demonstrated recently in [21] or in this manuscript. This characteristic makes our metasurfaces promising for the development of voltage-controlled tunable optical filters or modulators[25,26].

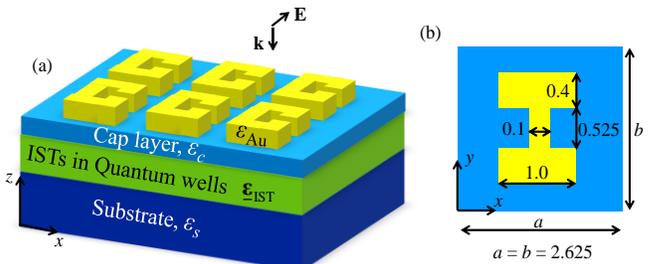

Fig. 1. (a) 3D view of a MS made of split ring resonators (SRRs) on top of a multilayered substrate comprising QWs for strong coupling purposes (dimensions are not in scale). The normal plane wave illumination, with electric field along *y*, is explicitly indicated. (b) Top view of the unit cell of a MS made of dogbone resonators. Dimensions in μm for the spatial scaling factor 1.0 are indicated (when scaling factor is varied, all the marked dimensions are scaled accordingly).

Strongly coupled systems using ISTs have been treated quantum-mechanically [7, 27, 28]. Moreover, it was shown that different resonator geometries may lead to different Rabi splittings [29], although an explanation to this phenomenon has not yet been provided. In this paper we observe instead that the near fields responsible for the strong coupling can be simply and accurately described by the near fields of electrostatic dipoles, an approach not considered before. Our analysis addresses and explains in a phenomenological manner the Rabi splitting dependence on the resonator geometry. This allows us to develop a closed-form expression for the metasurface resonators' capacitance that explicitly shows the strong-coupling contribution. From this, we develop an equivalent circuit network model that qualitatively and quantitatively describes the strong coupling processes.

We show that, for the case of a flat MS made of dogbone resonators presented in Fig. 1(b), the results of the circuit network model are in good agreement with both full-wave simulations and experimental measurements. Hence, our network represents a low order pole-zero polynomial expansion of the electromagnetic coupling between the MS resonance and the ISTs. Next, we use the circuit network model to analyze strong coupling for three resonator geometries and observe that the magnitude of the Rabi splitting can be increased by increasing the MS resonators' capacitance. This capacitance can be understood either as one coefficient of the polynomial expansion of the coupling or as the physical capacitance of the quasi-static model developed next.

## II. ELECTRODYNAMIC MODEL FOR FIELD-MATTER INTERACTION

Consider the structure in Fig. 1, where a MS of gold resonators (100 nm thick) is placed on top of a multilayered substrate comprising: an Al$_{.48}$In$_{.52}$As cap layer (30 nm thick) with relative permittivity $\varepsilon_c = 10.23$; a slab containing 20 repeat units of an In$_{.53}$Ga$_{.47}$As/Al$_{.48}$In$_{.52}$As heterostructure (12.5/20 nm) supporting ISTs at a frequency of $\approx 24.2$ THz; and an InP substrate with $\varepsilon_s = 9.3$. The gold permittivity $\varepsilon_{\text{Au}}$ is described using a Drude model [30] with parameters extracted from spectral ellipsometry measurements of a 100 nm thin gold film which yield a plasma angular frequency of $2\pi \times 2060 \times 10^{12}$ rad/s and a damping rate of $2\pi \times 10.9 \times 10^{12}$ 1/s. The ISTs are represented by anisotropic Lorentzian oscillators matched to experimental data, whose (relative) dielectric tensor is given as $\underline{\varepsilon}_{\text{IST}} = \varepsilon_t (\hat{\mathbf{x}}\hat{\mathbf{x}} + \hat{\mathbf{y}}\hat{\mathbf{y}}) + \varepsilon_z \hat{\mathbf{z}}\hat{\mathbf{z}}$ [20, 31, 32], with $\varepsilon_t = 10.97$ and

$$\varepsilon_z = \varepsilon_t + \frac{f_z \omega_0^2}{\omega_0^2 - \omega^2 - 2i\omega\gamma} \tag{1}$$

where $\omega_0 = 2\pi \times 24.2 \times 10^{12}$ [rad/s] is the IST angular frequency; $\gamma = 2\pi \times 10^{12}$ [1/s] represents the IST damping; and $f_z = 1.2$ is proportional to the IST oscillator strength, the doping density and intersubband matrix elements as described in [32]. [The monochromatic time harmonic convention, $\exp(-i\omega t)$, is used here and throughout the paper, and is suppressed hereafter.] Moreover, in our QW structure, the optically active transition happens between the ground state and the first excited state; the use of transitions between higher states inside the same QW, as for example demonstrated in [18], would lead to a wider Rabi splitting because of a larger oscillator strength.

Equation (1) together with the tensor $\underline{\varepsilon}_{\text{IST}}$ reveals the selection rule for optical excitation of ISTs: only light polarized along the QWs growth direction (here the $z$ direction) can promote electrons between different subbands.

For a MS array with sufficiently sub-wavelength elements (Fig. 1) illuminated at normal incidence, the propagating incident, reflected, and transmitted waves do not contain $z$-polarized fields. However, substantial $z$-polarized electric fields are generated in the near-field of the resonators within the array (see Appendix for more details). Due to the sub-wavelength dimensions of the MS resonators, the electric and magnetic near fields exhibit quasi-static behavior and, hence, are decoupled. Furthermore, based on the fact that the electric near field of a dynamic dipole is identically equal to the field of a *static* dipole [33], we model each resonator as a distributed set of electrostatic charges to capture the contribution of the $z$-polarized fields of the MS sub-wavelength resonators. Hence, to begin, we consider the *electrostatic* problem of a single dipole located on the free space side and at a short distance from the interface between free space and an anisotropic material with $\underline{\varepsilon}_{\text{IST}}$ (see Appendix for more details). Following the steps outlined in [34, 35], we estimate the static electric potential $\phi_e$ of the single dipole at any location in space. Assuming that the charges composing the single dipole are distributed over a tiny spherical surface with radius much smaller than the charge separation, we obtain an approximate formula for the capacitance associated to such a dipole located in very close proximity of the interface. Using superposition of effects, summing over the distributed set of charges on the flat resonator allows us to obtain an expression for the total capacitance $C$ of the resonator

$$C = \frac{\sqrt{\varepsilon_t \varepsilon_z} + 1}{\varepsilon_t + 1} C_{\text{ms}} = C_{\text{ms}} + C_{\text{eq}}^{\text{IST}} = \xi C_{\text{ms}} \tag{2}$$

where $C_{\text{ms}}$ is the MS resonators' capacitance per unit cell when no QWs are considered (i.e., $\varepsilon_z = \varepsilon_t$), $C_{\text{eq}}^{\text{IST}} = C_{\text{ms}} \left( \sqrt{\varepsilon_t \varepsilon_z} - \varepsilon_t \right) / (\varepsilon_t + 1)$ is the excess capacitance representing the strong coupling to the ISTs, and $\xi = \left( \sqrt{\varepsilon_t \varepsilon_z} - \varepsilon_t \right) / (\varepsilon_t + 1)$ is a "coupling coefficient". Strictly speaking, $\xi$ takes into account the perturbation of the electric field lines of the capacitor due to the presence of the ISTs in QWs with an anisotropic, frequency dependent dielectric constant $\varepsilon_z$ in Eq. (1). However, since the parameter $\xi$ is introduced only in presence of QWs, it can also be understood as a coupling coefficient. Note that we do not need to know the exact distribution of point charges in the MS since $C_{ms}$ will be determined using the method described below. Nonetheless, the linear nature of the sums required for the calculation of $C_{ms}$ and the establishment of an excess capacitance show that this method is general for any subwavelength resonator shape.

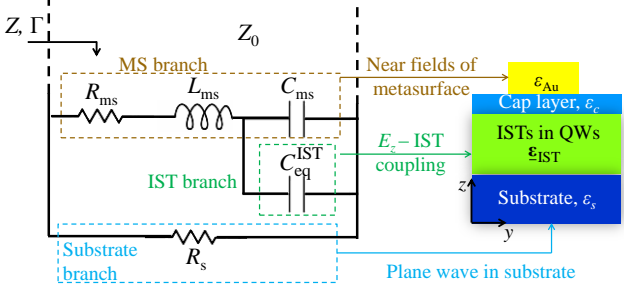

Fig. 2. Circuit network model with metasurface impedance $Z_{ms}$ composed by $R_{ms}$, $L_{ms}$, and $C_{ms}$, and the equivalent capacitor $C_{eq}^{IST}$. The correlation of each circuit element to the real structure is marked by the dashed contours.

Circuit network theory is a common way to model metamaterial properties [36-42]. Equation (2) together with the structure depicted in Fig. 1 allow for the construction of the equivalent circuit network model shown in Fig. 2, which embodies the physical processes of the bare flat MS as well as the strong coupling to the QWs. The substrate branch (light blue dashed contour) consists of a resistor $R_s$ that models the wave impedance of the dielectric layers below the MS itself. The MS is modeled via a series RLC resonant circuit [37] (brown dashed contour) to capture the MS resonant features (i.e., the "cavity" resonance). The coupling to the ISTs in QWs is represented by the complex valued equivalent excess capacitor $C_{eq}^{IST}$ which is arranged in parallel (green dashed contour) with the MS resonators' capacitance $C_{ms}$ according to Eq. (2). Therefore, when the resonators reside on an isotropic substrate (i.e., no QWs), $\varepsilon_z = \varepsilon_t$, $C_{eq}^{IST} = 0$, and the total capacitance is $C = C_{ms}$. When instead we consider the flat resonators on top of the QWs, the total capacitance is $C = C_{ms} + C_{eq}^{IST} = \xi C_{ms}$ as in Eq. (2). From the above equation we can estimate the coupling coefficient based on this quasi-static approximation as $\xi = \left(\sqrt{\varepsilon_t \varepsilon_z} + 1\right)/(\varepsilon_t + 1)$, which, notably, as a first approximation does not depend on the MS geometry but only on the QW design.

To obtain the circuit parameters, a full-wave simulation (using a commercial frequency-domain finite-element simulator [43]) of a MS without QWs is performed to determine the magnitude and phase of the plane wave reflection coefficient $\Gamma$ under normal incidence illumination with electric field polarized along $y$ (we refer the phase to the top metal surface). We then use a commercial circuit simulator [44] to match the simulated $\Gamma$ to the circuit reflection coefficient $\Gamma = (Z - Z_0)/(Z + Z_0)$, with $Z_0 = 377\ \Omega$ (illumination is from free space) and $Z$ the input impedance of the circuit in Fig. 2, evaluated just above the MS. This matching procedure in the frequency band 15 – 40 THz leads to the frequency-independent circuit parameters $R_s$, $R_{ms}$, $L_{ms}$, and $C_{ms}$. The value of $C_{eq}^{IST}$ is then simply obtained using Eq. (2) – *no further fitting is performed*. The circuit parameters for the dogbone resonator shown in Fig. 1(b) are given in Table 1.

### III. TOTAL ELECTRIC FIELD ENERGY EXCESS

To test the accuracy of our circuit model, we compare the energy and spectral properties obtained from the model with both full-wave simulations and experimental results. We first consider the electric energy. In a medium characterized by a dispersive, absorptive permittivity tensor $\underline{\varepsilon}_{IST}$, the time-averaged electric energy $W_e$ contained within the unit cell volume $V$ is evaluated as [45-47]

$$W_e = \int_V \left[\frac{1}{4}\varepsilon_0 \varepsilon_t |\mathbf{E}_t|^2 + \frac{1}{4}\varepsilon_0 \left(\varepsilon_z' + \frac{2\omega \varepsilon_z''}{2\gamma}\right)|E_z|^2\right] dV , \quad (3)$$

where $\varepsilon_z = \varepsilon_z' + i\varepsilon_z''$, and $\mathbf{E}_t$ and $E_z$ are the transverse and longitudinal components of the electric field, respectively, generated by a (monochromatic) normal plane wave with incident power $P = 1\ \text{W}$ per unit cell with area $A = ab$. Using the full-wave simulator, we perform the volumetric field integral over $V$ representing the various regions of space in the simulation setup of Fig. 1 as described next. We evaluate energies in two situations, with and without QWs. Then, we estimate the amount of energy excess $\Delta W_e$ (shown as the black curve in Fig. 3) due to the presence of the Lorentzian IST resonance in the QWs as

$$\Delta W_e = W_e^{QW} - W_e^{no\ QW}, \quad (4)$$

where $W_e^{QW}$ and $W_e^{no\ QW}$ are the total electric energies in the simulation with and without QWs, respectively, computed as follows. The term $W_e^{QW}$ is computed expanding the domain of integration into subdomains as

$$\begin{aligned}W_e^{QW} = &\frac{1}{4}\varepsilon_0\left(\varepsilon_z' + \frac{2\omega\varepsilon_z''}{2\gamma}\right)\int_{V_{QW}} \left|E_z^{QW}\right|^2 dV \\ &+ \frac{1}{4}\varepsilon_0 \varepsilon_t \int_{V_{QW}} \left|\mathbf{E}_t^{QW}\right|^2 dV + \sum_i \frac{1}{4}\varepsilon_0 \varepsilon_i \int_{V_i} \left|\mathbf{E}^{QW}\right|^2 dV\end{aligned}, \quad (5)$$

where the first two terms refer to the QW region in the unit cell and in the last term $i$ = {vacuum (200 nm), cap layer} represents the two volumetric regions outside the QWs, in the unit cell. Note that as a first approximation we include only a narrow portion (200 nm thickness) of the vacuum region as we only account for the near-field energy in close proximity of the resonators; likewise, we do not include the InP substrate as near fields decay rapidly within the QWs. Note that the permittivity $\varepsilon_z$ is dispersive in the QWs due to the ISTs and therefore the energy must be calculated accordingly. The energy term $W_e^{\text{no QW}}$ in absence of the QW (and thus without IST) is instead computed as

$$W_e^{\text{no QW}} = \sum_j \frac{1}{4} \varepsilon_0 \varepsilon_j \int_{V_j} \left| \mathbf{E}^{\text{no QW}} \right|^2 dV, \quad (6)$$

with $j$ = {vacuum (200 nm), cap layer, isotropic QW layer}, where the permittivity is non-dispersive in all three volumetric regions. Approximations to the volumetric regions mentioned in regards to Eq. (5) are applied also to Eq. (6). These energies are computed numerically using total fields (i.e., including both plane wave and reactive components) obtained from full-wave simulation, for both cases of absence and presence of the QWs, properly dividing the integration regions and polarizations and considering the IST-induced frequency dispersion in the QWs. In close proximity of the resonators (i.e., near field) the total field is as a first approximation dominated by the reactive components.

We now show that $\Delta W_e$ per unit cell can also be determined using the circuit model, and compare it to the full-wave result. We can evaluate the magnitude of the driving (incident) electric field as $E_0 = \sqrt{2Z_0 P / A}$. The total transverse electric field at the input impedance terminals is then given by $\mathbf{E}_t = E_0 (1 + \Gamma) \hat{\mathbf{y}}$. The term $W_e^{\text{QW}}$ is then computed assuming an incident traveling voltage wave of $V_{\text{inc}} = E_0 b$ as

$$W_e^{\text{QW}} = \frac{1}{4} \left( C' + \frac{2\omega C''}{2\gamma} \right) \left| V_C^{\text{QW}} \right|^2 \quad (7)$$

where $V_C^{\text{QW}}$ (evaluated as shown in the Appendix) is the voltage acting on the total capacitor $C = C' + iC''$ given by Eq. (2). In Eq. (7), we have applied the same dispersive condition as in Eq. (3) in the framework of circuit theory [48], assuming a damping as in Eq. (1). The term $W_e^{\text{no QW}}$ is instead computed simply as

$$W_e^{\text{no QW}} = \frac{1}{4} C_{\text{ms}} \left| V_C^{\text{no QW}} \right|^2 \quad (8)$$

where $V_C^{\text{no QW}}$ is the voltage acting on the MS resonators' capacitor $C_{\text{ms}}$ (evaluated as shown in the Appendix). The energy difference due to the QWs obtained using this procedure is shown as the red dashed curve in Fig. 3. The agreement between the energy $\Delta W_e$ evaluated via full-wave simulations and circuit network theory shown is quite remarkable, given that the circuit network model is based on a quasi-static approximation and the coupling coefficient $\xi$ in closed form as in Eq. (2), and hence is extremely simple. Note that if instead of using Eq. (2) we evaluate $C_{\text{eq}}^{\text{IST}}$ via a second fitting procedure using the circuit simulator similar to what was done for $C_{\text{ms}}$ but this time using full-wave data in the presence of QWs, the agreement for the peaks locations would be further improved, validating the circuit topology in Fig. 2; however our point is that even the simple formula in Eq. (2), with coupling coefficient $\xi$ determined analytically, provides a good explanation.

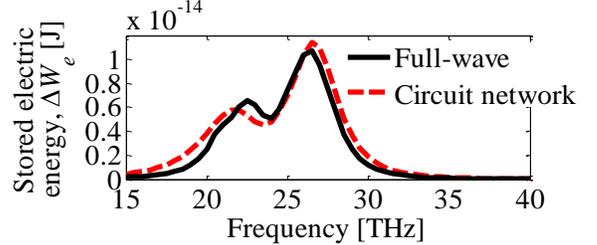

Fig. 3. Stored electric energy $\Delta W_e$ per unit cell computed by full-wave simulations (solid black) and circuit network model (red dashed) for the dogbone resonator case in Fig. 1(b).

## IV. SPECTRAL REFLECTIVITY

Figure 4(a) shows the $|\Gamma|^2$ reflectivity maps obtained from full-wave simulations of the MS of dogbone resonators on top of QWs as a function of the bare cavity resonance, controlled by scaling the MS geometry. To generate this map, we simulated a set of metasurfaces for which all the spatial dimensions of the MS of dogbone resonators are scaled by a common scaling factor that varied between 0.7 and 1.2 [relative to the dimensions shown in Fig. 1(b)]. For each scaling factor, we performed a full-wave simulation of the MS without the QWs to determine the bare cavity resonance frequency as the location of maximum reflectivity. Next we performed simulations including the QWs and plotted the obtained reflection spectra versus the bare cavity frequency. Figure 4(b) shows the corresponding maps obtained from the circuit model. For each scaling factor used in Fig. 4(a), we use the full-wave simulation result (without the QWs) to obtain the circuit parameters $R_s$, $R_{\text{ms}}$, $L_{\text{ms}}$, and $C_{\text{ms}}$. We then use Eq. (2) to calculate $C_{\text{eq}}^{\text{IST}}$ which is then inserted in the circuit model to obtain the reflection spectra. Finally, Fig. 4(c) shows the experimentally measured reflectivity spectra for a set of metasurfaces with the same scaling factors [21]. The splitting around 24 THz is clearly visible in the experiment as well as in the simulations. In addition, extremely good agreement between the circuit model and the full-wave simulation is observed. The small reflectivity magnitude mismatch in the experimental reflectivity with respect to other results is attributed to fabrication imperfections. The results in Fig. 4 together with those in Fig. 3 show that Eq. (2) is correct and confirm the validity of the circuit model introduced in Fig. 2. In turn, this implies that the electrostatic near fields of dipoles can be used to accurately describe the strong coupling processes, a result that may not be straightforward in complex systems as the one here analyzed. To the authors' knowledge, such an approach has never been described before.

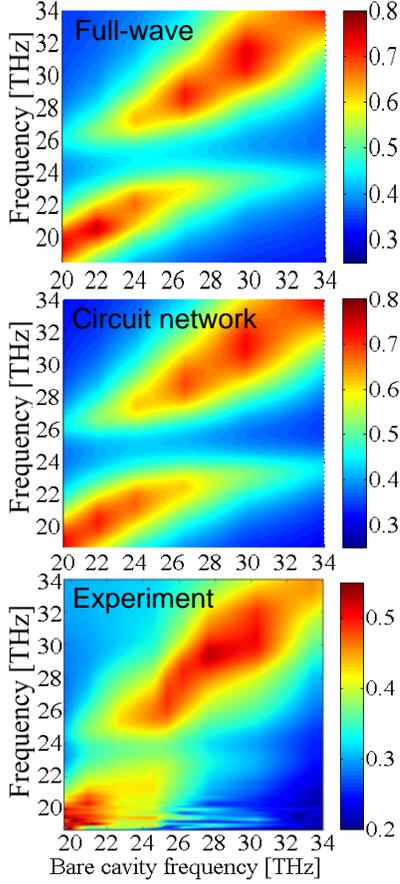

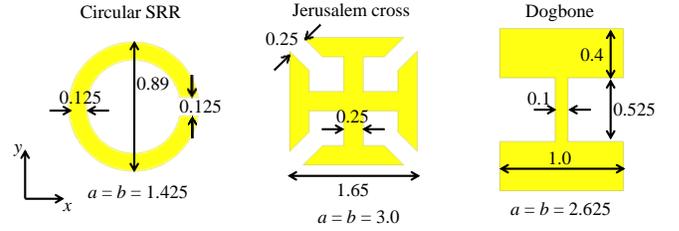

Fig. 5. Three resonator structures considered, along with dimensions (in μm) for a scaling factor of 1.0.

To elucidate the effect of the MS elements on strong coupling, we consider now the three MSs based on the three different resonator geometries shown in Fig. 5 (dimensions of structures for scaling factor 1.0 are indicated therein): (i) circular SRR; (ii) Jerusalem cross; and (iii) dogbone. The three MSs are on top of the same QWs described by $\underline{\varepsilon}_{IST}$. For each resonator type, we perform both full-wave and circuit model simulations over a range of scaling factors to map out the polariton branches, corresponding to the locations of two reflectivity maxima. The magnitude and phase of the reflection coefficient $\Gamma$ for the three resonator geometries, for the scaling factors that bring the IST and bare cavity resonance into near coincidence, are shown in Fig. 6. The top row of Fig. 6 shows the case with no QWs. In this case, a single, primary resonance is observed for the three geometries, and in the vicinity of this resonance the circuit model (dashed curves) is in good agreement with full-wave results (solid curves). This result further shows that our fitting procedure recovers circuit parameters that accurately reproduce the resonant features of the MS when $C_{eq}^{IST}=0$. Inserting $C_{eq}^{IST}$, obtained from Eq. (2), in the circuit model corresponds to the case where QWs are present, and leads to a splitting of the primary resonance as is shown in the bottom row of Fig. 6. Once again, good agreement between the circuit model and the full-wave results is observed in the vicinity of the resonances. We recall that circuit parameters are retrieved considering the frequency range 15-40 THz, and that if one desires to match also higher frequencies, more complicated circuit branches will be necessary.

Figure 7 shows the two polariton branches as a function of the bare cavity frequency (varied changing scaling factor) for the three resonators obtained using the circuit model and full-wave simulation: excellent agreement with the full-wave results is obtained. This plot nicely shows the splitting due to strong coupling and, through the use of Eq. (6) in [8], we are able to estimate the Rabi splitting $\Omega_R$ which is reported in Table 1. Table 1 also lists the circuit parameters $R_s$, $R_{ms}$, $L_{ms}$, and $C_{ms}$ for each resonator geometry for the scaling factors that lead to nearly coincident MS and ISTs resonant frequencies. Specifically, these parameter correspond to the dogbone resonator and circular SRR resonator with a scaling of 1.0, and the Jerusalem cross resonator with a scaling of 0.9. Note that in the parameter retrieval we have allowed $R_s$ to slightly change for the different structures because higher

Fig. 4. Maps of reflectivity $|\Gamma|^2$ from full-wave simulation, circuit network model, and experimental setup for the dogbone structure in Fig. 1(b) with QWs. Only the values relative to bare cavity frequencies pertaining to scaling factors 0.7, 0.8, 0.9, 1.0, 1.1, and 1.2 were used. Continuous plots are shown for better visibility.

## V. PARAMETERS CONTROLLING THE COUPLING STRENGTH

It follows from Eq. (2) that only two parameters affect strong coupling: the capacitance $C_{ms}$ and the coupling coefficient $\xi$. The coupling coefficient $\xi$ is the only parameter that contains the Lorentzian dispersion in ISTs through the quantity $\sqrt{\varepsilon_t \varepsilon_z}$, and does not contain information on the MS element shape, further stressing the elegance and simplicity of the proposed model. This means that when we compare the performance of different resonator geometries using the same QWs, $\xi$ will be unchanged. [We remark that more efficient ISTs in QWs (i.e., stronger anisotropy, and thus larger $\xi$) might lead to larger splitting.] Therefore, differences in the magnitude of the coupling obtained with different resonator geometries will only arise if $C_{ms}$ changes. Other MS circuit parameters will, in principle, also affect the spectral properties as they contribute to reflection magnitude and resonance quality factor.

order TM polarized modes may propagate into the substrate for the considered unit cell sizes and thus contribute to extra losses in the structure.

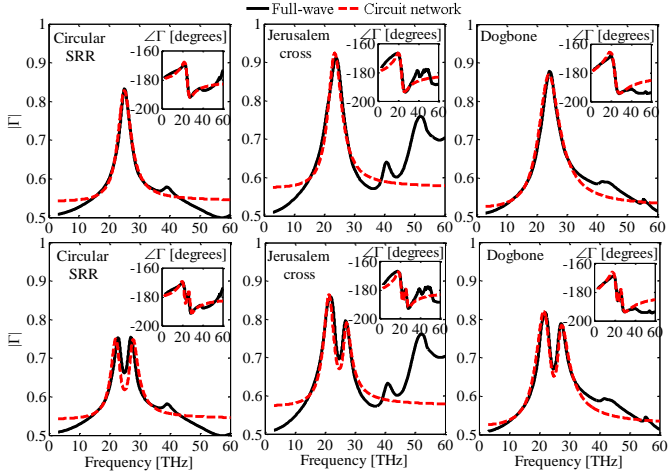

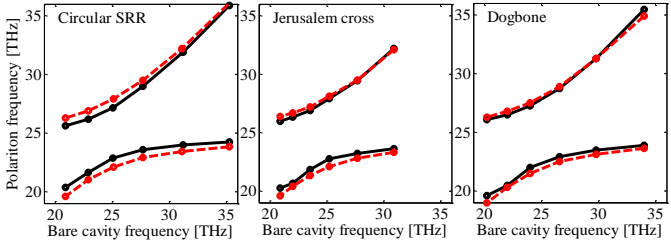

Fig. 6. Magnitude and (unwrapped) phase of $\Gamma$ from full-wave simulation (black solid) and circuit network model (red dashed) for the three structures in Fig. 5 without QWs (top row) and with QWs (bottom row).

Fig. 7. Plot of the two polariton branches versus the bare cavity frequency for the three resonator structures in Fig. 5 obtained from full-wave simulation (black solid) and circuit network theory (red dashed).

Importantly, we see from Table 1 that the capacitance values of the dogbone and Jerusalem cross resonators are similar in magnitude, leading to similar Rabi splittings. In contrast, the capacitance of the circular SRR resonator is smaller, causing the narrower splitting observed in the full-wave result in the bottom row of Fig. 6 and in Fig. 7. This confirms the prediction made earlier that the splitting only depends on the value of $C_{\text{ms}}$, assuming the coupling coefficient $\xi$ does not change for the three cases. To further confirm this observation, we re-designed the circular SRR with wider traces to increase its capacitance (diameter of 1.15 μm and traces of 0.35 μm for scaling factor 1.0, exhibiting $C_{\text{ms}} \approx 13.7$ aF). With such resonator, we get an even larger Rabi splitting of $\Omega_R/(2\pi) \approx 2.4$ THz. Using our theory we can infer that microcavity designs comparable to the ones presented in [49, 50] could be an excellent choice to enhance light-matter coupling in presence of a metallic backplane.

We have seen that the Rabi splitting directly depends on the capacitance $C_{\text{ms}}$ and thus on the resonator physical parameters if approximate expressions of the capacitance for the considered resonator geometries are known. For isolated SRRs without substrate, the capacitance can be expressed as [51] $C_{\text{SRR}} = C_{\text{gap}} + C_{\text{surf}}$, with $C_{\text{gap}} = \varepsilon_0 hw/g + \varepsilon_0(h+w+g)$ (which comprises the parallel-plate capacitor formed by the gap and a correcting term due to the fringing fields) and $C_{\text{surf}} = \left[2\varepsilon_0(h+w)\log(4R/g)\right]/\pi$ the surface capacitance, with $h$ the SRR thickness, $w$ the SRR trace width, $g$ the SRR gap, and $R$ the SRR inner radius. Using such formulas for the two SRR geometries analyzed above, we obtain $C_{\text{SRR}} \cong 7$ aF and $C_{\text{SRR}} \cong 12.6$ aF, respectively, showing the same order of magnitude of the results reported above for the two metasurfaces. More accurate formulas would require the account of the substrate. However, similar trends with physical parameters are expected.

Table 1. Rabi splitting from full-wave simulations and circuit parameter values in Fig. 2 for the three resonator shapes in Fig. 5 without QWs.

| Resonator | $\Omega_R/(2\pi)$ [THz] | $C_{\text{ms}}$ [aF] | $L_{\text{ms}}$ [pH] | $R_{\text{ms}}$ [$\Omega$] | $R_s$ [$\Omega$] |
|---|---|---|---|---|---|
| Circular SRR | 2.1 | 8.7 | 4.6 | 51.8 | 111.7 |
| Jerusalem cross | 2.6 | 14.6 | 3.2 | 17.6 | 101.9 |
| Dogbone | 2.9 | 16.7 | 2.6 | 34.2 | 117.4 |

## VI. CONCLUSION

We have reported an equivalent electrodynamic model for strong coupling between a MS and ISTs in QWs, by resorting to an equivalent circuit network theory and the electrostatic near-fields of the flat MS dipoles. Such a model qualitatively and quantitatively explains the energy stored near the MS and the energy exchanged between the MS and ISTs in QWs. It is also able to recover the reflection and transmission spectral properties of systems where electromagnetic fields and ISTs in QWs are strongly coupled. We have adopted this model at mid-infrared frequencies, but our findings will apply in any given wavelength range. Despite the three dimensional (volumetric) nature of the light-matter coupling, we observed that the strong coupling mechanism is governed mainly by only two parameters: the coupling parameter $\xi$ which depends on the QW design and not on the MS geometry; and the MS resonators' capacitance $C_{\text{ms}}$ in absence of QWs. Although the use of the electrostatic fields of a dipole to describe the strong coupling process may at first seem surprising, its validity is based on two main observations. First, the QWs interact only with $z$-polarized electric fields. Due to the sub-wavelength period of the array, the propagating transmitted and reflected electric fields contain no $z$-component. Thus, $z$-polarized fields occur only in the quasi-static (i.e, near-field) zone of the MS resonators and are well described by electrostatic dipolar fields. Second, near-field interactions among resonators in the MS are negligible due to the rapid decay of the fields. Through the use of this parameterization we have demonstrated that the Rabi splitting increases with the capacitance $C_{\text{ms}}$, thereby providing a phenomenological explanation of why different Rabi splittings have been observed with different resonator shapes. The results reported in this manuscript provide us with a pathway to increase the Rabi splitting by: i) optimizing the metasurface by increasing the resonators' capacitance; and ii) optimizing the QW design (for example, by increasing the oscillator strength). Such optimization may enable us to go beyond strong coupling regime by using planar metasurfaces coupled to ISTs in QWs at infrared frequencies.

## ACKNOWLEDGMENT

This work was performed, in part, at the Center for Integrated Nanotechnologies, a U.S. Department of Energy, Office of Basic Energy Sciences user facility. Portions of this work were supported by the Laboratory Directed Research and Development program at Sandia National Laboratories. Sandia National Laboratories is a multi-program laboratory managed and operated by Sandia Corporation, a wholly owned subsidiary of Lockheed Martin Corporation, for the U.S. Department of Energy's National Nuclear Security Administration under contract DE-AC04-94AL85000.

S.C. and F.C. thank Ansys Inc. and AWR Corporation for providing HFSS and Microwave Office, respectively, which were instrumental in this work.

## APPENDIX

*1. Steps to achieve Eq. (2)*

Consider a uniaxial anisotropic material with (relative) dielectric tensor

$$\underline{\underline{\varepsilon}} = \varepsilon_t \left( \hat{\mathbf{x}}\hat{\mathbf{x}} + \hat{\mathbf{y}}\hat{\mathbf{y}} \right) + \varepsilon_z \hat{\mathbf{z}}\hat{\mathbf{z}}. \quad (9)$$

The displacement field is expressed as $\mathbf{D} = \underline{\underline{\varepsilon}} \cdot \mathbf{E}$. We also know that the equation $\nabla \cdot \mathbf{D} = \rho_e$ has to be satisfied, where $\rho_e$ is the electric charge density. Limiting the analysis to electrostatic fields because the structures used in the MS are sub-wavelength, the electric field in proximity of the MS resonators is well approximated by

$$\mathbf{E} = -\nabla \phi_e \quad (10)$$

where $\phi_e$ is the scalar electric potential generated by the charge density $\rho_e$ accumulated at the ends of the dogbone shape in Fig. 1(b), for example.

Consider now the scenario of a point charge $\rho_e = q_e \delta(\mathbf{r} - \mathbf{r}_e)$, with $\mathbf{r}_e = s\hat{\mathbf{z}}$, located close to the interface between free space and a uniaxial anisotropic material (the ISTs in QWs) as shown in Fig. 8, where the point charge is in the free space side at a distance $s$ from the interface.

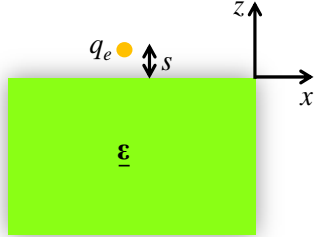

Fig. 8. Point charge (in the free space region) located near the interface between an isotropic and a uniaxial anisotropic material.

Following the steps in [34, 35], the potential $\phi_e$ at a point $(x,y,z)$ within free space is given by

$$\phi_e = \frac{q_e}{4\pi\varepsilon_0} \left[ \frac{1}{R_-} - \frac{\frac{\sqrt{\varepsilon_t \varepsilon_z} - 1}{\sqrt{\varepsilon_t \varepsilon_z} + 1}}{R_+} \right] \quad (11)$$

where $R_\pm = \sqrt{x^2 + y^2 + (z \pm s)^2}$ represents the distance either from the charge or from its image, whereas within the anisotropic material is

$$\phi_e = \frac{1}{4\pi\varepsilon_0} \frac{2q_e}{\sqrt{\varepsilon_t \varepsilon_z} + 1} \frac{1}{\sqrt{x^2 + y^2 + \left(\sqrt{\frac{\varepsilon_t}{\varepsilon_z}} z - s\right)^2}}. \quad (12)$$

Note that the potential is continuous across the interface at $z = 0$. Knowing Eqs. (11)-(12) for a single charge, we now consider a transverse dipole made of two displaced and opposite charges $\pm q_e$ with separation $d$ located very close to the interface, i.e., $s = 0^+$. We further assume that the charges are uniformly distributed around small spherical surfaces with radius $r_c \ll d$ (so that we can consider it uniformly distributed). We can then estimate the capacitance associated with such a two-charge system as

$$C = \frac{q_e}{\phi_+ - \phi_-}. \quad (13)$$

where $\phi_+ - \phi_-$ is the potential difference between the two charges, $\phi_+$ and $\phi_-$ are the potentials at the surface of the positive and negative charge spherical distributions, respectively.

Using Eq. (11), the potential at any observation point within free space is

$$\phi_e = \frac{q_e}{4\pi\varepsilon_0 r_1} \left[ 1 - \frac{\sqrt{\varepsilon_t \varepsilon_z} - 1}{\sqrt{\varepsilon_t \varepsilon_z} + 1} \right] - \frac{q_e}{4\pi\varepsilon_0 r_2} \left[ 1 - \frac{\sqrt{\varepsilon_t \varepsilon_z} - 1}{\sqrt{\varepsilon_t \varepsilon_z} + 1} \right] \quad (14)$$

where $r_1$ and $r_2$ are the distances from the given observation point to the positive and negative charge positions, respectively. In particular, for the observation point at the surface location of the positive distributed charge, we have $r_1 = r_c$ and $r_2 \approx d$ because $r_c \ll d$. Therefore, the potential $\phi_+$ is given by

$$\phi_+ = \frac{q_e}{4\pi\varepsilon_0} \frac{2}{\sqrt{\varepsilon_t \varepsilon_z} + 1} \left( \frac{1}{r_c} - \frac{1}{d} \right). \quad (15)$$

Similarly, at the surface location of the negative distributed charge, we have $r_1 \approx d$ because $r_c \ll d$ and $r_2 = r_c$, so the potential $\phi_-$ is given by

$$\phi_- = \frac{q_e}{4\pi\varepsilon_0} \frac{2}{\sqrt{\varepsilon_t \varepsilon_z}+1}\left(\frac{1}{d}-\frac{1}{r_c}\right). \qquad (16)$$

Thus the potential difference is equal to

$$\phi_+ - \phi_- = \frac{q_e}{4\pi\varepsilon_0} \frac{4}{\sqrt{\varepsilon_t \varepsilon_z}+1}\left(\frac{1}{r_c}-\frac{1}{d}\right). \qquad (17)$$

Therefore, we can obtain the formula for the capacitance, under the assumption $r_c \ll d$:

$$C = \pi\varepsilon_0 r_c \left(\sqrt{\varepsilon_t \varepsilon_z}+1\right). \qquad (18)$$

Note that in this construction, the charge size $r_c$ does not need to be known. [We note that any distribution can be used, which would however result in different geometrical factors in Eq. (18) instead of the factor $r_c$.] The capacitance associated to the same two-charge system on top of a fully isotropic material (i.e., when the anisotropic layer is replaced by an isotropic material with $\varepsilon_z = \varepsilon_t$) is $C_0 = \pi\varepsilon_0 r_c (\varepsilon_t + 1)$. (In free space, $\varepsilon_t = 1$ and $C_0 = 2\pi\varepsilon_0 r_c$ as the result in literature for a two-charge system in free space [52-56].) Therefore, we can rewrite Eq. (18) as

$$C = \frac{\sqrt{\varepsilon_t \varepsilon_z}+1}{\varepsilon_t + 1} C_0 = \xi C_0 \qquad (19)$$

and most importantly, the ratio $C/C_0 = \left(\sqrt{\varepsilon_t \varepsilon_z}+1\right)/(\varepsilon_t+1)$ does not depend on the unknown size $a$ of the charge distribution.

Note that a flat resonator (as those in Fig. 1 and Fig. 5) is in general described by a distributed set of charge pairs, i.e., dipoles sharing the same potential difference under the static approximation. To determine the same functional dependence $C/C_0$ shown above also for a set of charges, we assume that the charge distribution is flat, i.e., all charges are at the same distance from the substrate underneath. This in turn means that the effect of each pair of charges is accounted for by Eq. (19), and after applying the superposition of effects Eq. (19) can be generalized to Eq. (2) by substituting $C_{ms}$ for $C_0$, which results from the summation over the resonator's distributed charge distributions. Note that this is an approximate equation to model the complex system of a flat MS and the ISTs in QWs underneath. The use of Eq. (19) provides us with the means of understanding which parameters affect the strong coupling processes and is found to be remarkably accurate for the analyzed flat metasurfaces, also based on what shown in the next appendix.

*2. Comparison of electric fields under a MS made of dogbone resonators, calculated with electrostatic and full-wave models*

We show that the field generated by a set of electrostatic charges placed on the paddles of a dogbone resonator as in Fig. 9 recovers the electric field calculated via full-wave simulations, validating the quasi-static approximation used in the circuit model.

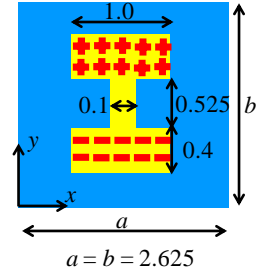

$a = b = 2.625$

Fig. 9. Sketch of the two sets of electrostatic positive and negative charges displaced on the paddles of a dogbone resonator (the charges are uniformly distributed). Dimensions in μm are reported.

To do so, we perform two full-wave simulations of a MS made of dogbone resonators: (i) as in Fig. 1 where the anisotropic region has finite thickness, denoted as Full-wave 1; (ii) as in Fig. 8 where the anisotropic region has infinite thickness, denoted as Full-wave 2. In both simulations, the environment below the resonators starts at $z = 0$. We then apply the formalism reported in the Appendix 1, Eq. (10), at the IST resonance frequency of 24.2 THz and evaluate the electrostatic field generated by two sets of opposite charges uniformly distributed on the two paddles of the dogbone as in Fig. 9 (we consider them to be at $z = s = 50$ nm, i.e., midway of the thickness of the dogbone resonator). The anisotropic region starts at $z = 0$.

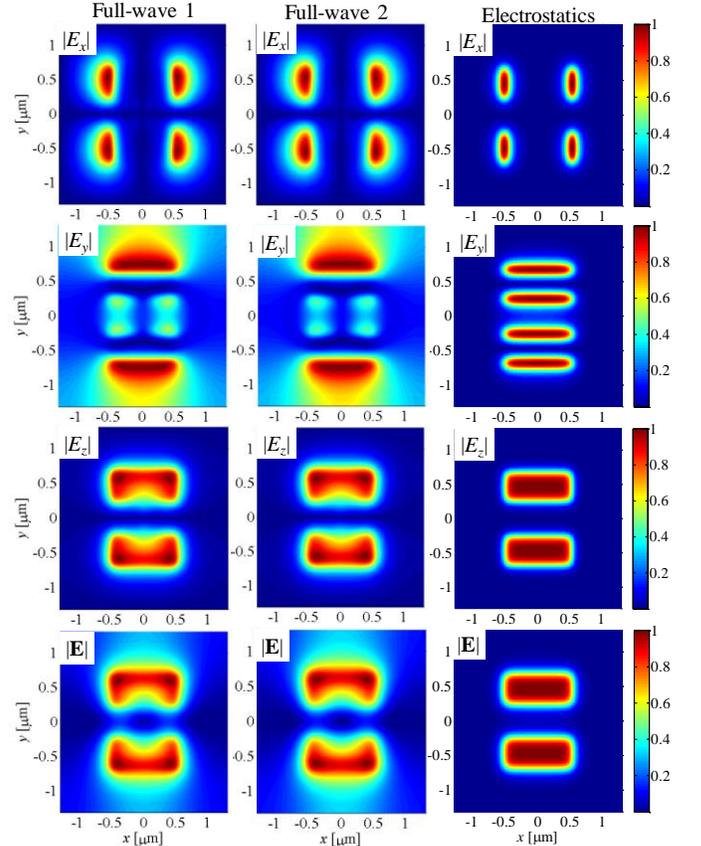

Fig. 10. Magnitude of the electric field components in the *x*-*y* plane (cut at $z = -100$ nm for all simulation setups, within the anisotropic region) in the case of MS made of dogbone resonators. Comparison between electrostatic

and full-wave solutions shows agreement. The difference between the two full-wave simulations is in the anisotropic region thickness.

We then show in Fig. 10 the magnitude of the electric field components (each component is normalized to its own maximum) in the *x-y* plane (cut at $z = -100$ nm for all simulation setups, within the anisotropic region). One can notice that the two full-wave simulations provide very similar electric field patterns (also the decay along the *z* direction is similar, not shown for brevity), so the approximation that assumes the anisotropic medium below the MS to be half space can be safely applied to determine the electric fields and so the capacitance. We want to mention that the strongest near-field component at resonance (in presence of QWs) in full-wave simulations is $E_z$, the field component required for the excitation of the ISTs in QWs according to the dipole selection rule. This in turn shows that most of the electric energy in presence of QWs is stored by $E_z$ in the near fields of the resonators. (Note that for the fields generated by the metasurface *without QWs*, the electric energy would be approximately evenly split between $E_y$ and $E_z$.) We remark however that in the electrostatic case, which provides field distributions in remarkable agreement, we are simulating an isolated unit cell and not a periodic set of dogbones, therefore coupling among MS unit cells is neglected in this simulation. Moreover, we are considering a uniformly distributed set of charges on the dogbone paddles, and this seems not to be exactly the case when looking at full-wave simulation results. Nonetheless, the result in Fig. 10 proves that the near-zone electric field generated by the dogbones can be predicted by a distribution of electric dipoles, and as such the electrostatic approximation adopted in part of this paper is valid. Finally, note that in the equivalent circuit model the capacitors mainly store only electric energy, which is associated to the volumetric integrals integrating the strong energy density over the small volume above and below the MS elements. In particular $C_{\text{eq}}^{\text{IST}}$ is defined assuming only $E_z$ and since this field component cannot propagate away from the MS, it is totally defined in the near-field.

Though the circuit model uses capacitors to represent the MS-QW coupling, highlighting the $E_z$ electric energy channel for the strong coupling, the circuit model in Fig. 2 could be devised also based on a simple pole-zero expansion in the polynomial approximation of the numerator and denominator of the reflection coefficient $\Gamma$. Indeed, it is well known that certain polynomial expansions are associated with realizable (i.e., physical) passive circuit elements and the polynomial coefficients can be expressed in terms of lumped elements values. Therefore, despite the use of a static approximation in this paper, such equivalent circuit models can also be generalized to more complicated structures.

*3. Computation of $V_C^{\text{QW}}$ and $V_C^{\text{no QW}}$ in Eqs. (7) and (8)*

We provide here the steps to compute the voltages $V_C^{\text{QW}}$ and $V_C^{\text{no QW}}$ across the capacitors, with and without QWs in the region below the MS, required for the calculation of the energy $\Delta W_e$ via circuit network theory.

Given the circuit model in Fig. 2, and knowing the incident traveling voltage wave $V_{\text{inc}} = E_0 b$, evaluated just above the MS layer, the total transverse voltage at the input impedance terminals accounts also for the reflected wave as $V = E_0 (1 + \Gamma) b$. The current flowing in the upper branch of the circuit network can be calculated as follows. Let us consider first the case without QWs (i.e., $C_{\text{eq}}^{\text{IST}} = 0$). In such case, the current flowing in the MS branch is given by

$$I_{\text{ms}} = \frac{V}{Z_{\text{ms}}} = \frac{E_0 (1+\Gamma) b}{R_{\text{ms}} - i\omega L_{\text{ms}} + i/(\omega C_{\text{ms}})}. \quad (20)$$

The voltage across the MS capacitor is then given by the generalized Ohms's law for impedances:

$$V_C^{\text{no QW}} = \frac{i}{\omega C_{\text{ms}}} I_{\text{ms}}. \quad (21)$$

In the presence of QWs, similar steps can be applied. In particular assuming the two capacitors in parallel, $I_{\text{ms}}$ is now given by

$$I_{\text{ms}} = \frac{V}{Z'} = \frac{E_0 (1+\Gamma) b}{R_{\text{ms}} - i\omega L_{\text{ms}} + i/\left[\omega\left(C_{\text{ms}} + C_{\text{eq}}^{\text{IST}}\right)\right]}. \quad (22)$$

The voltage across the IST capacitor is then given by

$$V_C^{\text{QW}} = \frac{i}{\omega\left(C_{\text{ms}} + C_{\text{eq}}^{\text{IST}}\right)} I_{\text{ms}}. \quad (23)$$